\documentclass[a4paper,12pt]{article}
\usepackage{amsmath,amssymb,vmargin,command}
\usepackage[dvips]{graphicx}
\setmarginsrb{1in}{1in}{1in}{0.75in}{0in}{0in}{12pt}{0.5in}
\title{Propagation of Axi-Symmetric Nonlinear Shallow Water Waves
    over Slowly Varying Depth}
\author{S.~M.~Killen and R.~S.~Johnson}
\date{\footnotesize{\mbox{Department of Mathematics},
    \mbox{University of Newcastle Upon Tyne},\\ \mbox{Newcastle Upon
      Tyne}, \mbox{NE1 7RU}, \mbox{United Kingdom}}}
\begin{document}
\maketitle
\begin{abstract}
  A problem in nonlinear water-wave propagation on the surface of an
  inviscid, stationary fluid is presented.
  
  The primary surface wave, suitably initiated at some radius, is
  taken to be a slowly evolving nonlinear cylindrical wave (governed
  by an appropriate Korteweg-de Vries equation); the depth is assumed
  to be varying in a purely radial direction.
  
  We consider a $\sech^2$ profile at an initial radius (which
  is, following our scalings, rather large), and we describe the
  evolution as it propagates radially outwards.  This initial profile
  was chosen because its evolution over constant depth is understood
  both analytically and numerically, even though it is not an exact
  solitary-wave solution of the cylindrical KdV equation.  The
  propagation process will introduce reflected and re-reflected
  components which will also be described.  The precise nature of these
  reflections is fixed by the requirements of mass conservation.
  
  The asymptotic results presented describe the evolution of the
  primary wave, the development of an outward shelf and also an inward
  (reflected) shelf.  These results make use of specific depth
  variations (which were chosen to simplify the solution of the
  relevant equations), and mirror those obtained for the problem of
  1-D plane waves over variable depth, although the details here are more
  complex due to the axi-symmetry.
\end{abstract}


\section{Introduction}

The propagation of plane solitary waves over variable depth is now
well understood
(\cite{johnson73,miles79,knicker80,knicker85,johnson94}).  In this
problem it is shown that, starting from an initial solitary-wave
solution of the KdV equation (\cite{korteweg}), the propagation over a
region of varying depth introduces a shelf directly behind the primary
wave, a left-going (reflected) 'shelf' and another right-going 'shelf'
(re-reflection).  All four of these components are required in order
to ensure that the global $O(1)$ mass is conserved.

The study of the propagation of axi-symmetric (or cylindrical) waves
is not, however, so well understood.  These waves, governed by the
cylindrical KdV equation
\begin{displaymath}
  u_\eta + \frac{u}{2\eta} +uu_\xi + \frac{1}{2}u_{\xi\xi\xi}=0,
\end{displaymath}
(as first written in \cite{maxon74}, and used in their study of
ion-acoustic waves), have additional geometric complexities which
complicate their study.

To date, the work involving cylindrical solitons and the cylindrical
KdV equation has been concerned with finding analytic solutions
(\cite{calogero78,hirota79,johnson79,santini79,santini80,nakamura81})
or using numerical methods to investigate the evolution of given
initial profiles (\cite{maxon74}).  There have, however, been
investigations involving additional physical features, for example,
including an underlying shear flow \cite{johnson90}; but nothing to
the extent of those for the plane wave problem.

The aim of this work is to draw upon the ideas and techniques used in
the study of plane-wave propagation over variable depth, and apply
them to the axi-symmetric problem.  Some of the issues raised in
\cite{johnson99} will play an important r\^ole; in fact, because of
this, the $\sech^2$ initial profile was used instead of the
appropriate solitary-wave solution of the cylindrical KdV equation, in
order to highlight the correspondence.

\section{Governing Equations}
We begin by introducing the equations and boundary conditions which
will be used for this investigation. We make several assumptions about
the ambient state of the fluid, which is taken as stretching to
infinity in all horizontal directions, and lies between a free upper
surface and a solid impermeable lower surface.  We will consider an
inviscid fluid, which implies that there will be no viscous stresses
acting on the surface of the fluid, thus eliminating the possibility
of the surface disturbances being driven by winds, for example (and also
eliminating the existence of a boundary layer flow along the bottom).

The fluid will be initially stationary (in its unperturbed state).  In
addition we assume the fluid to be incompressible, and we neglect the
effects of surface tension.

\begin{figure}
  \begin{center}
    \includegraphics[height=7cm]{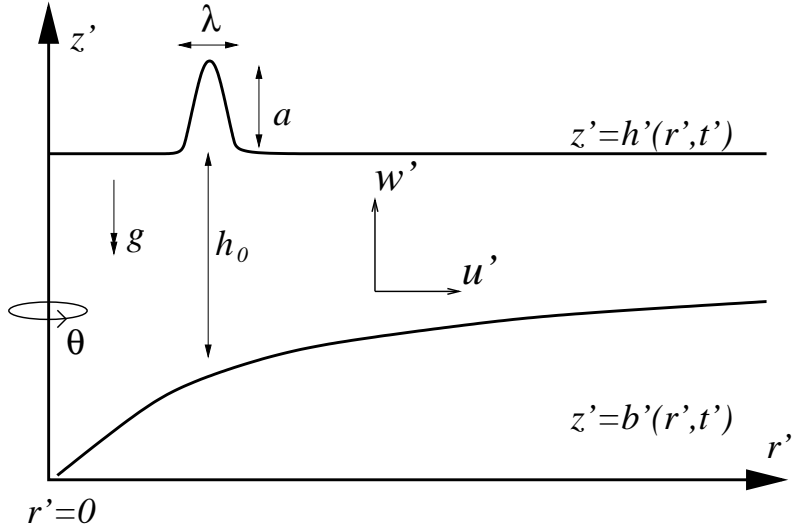}
  \end{center}
\caption{Configuration of the problem}\label{fig:config}
\end{figure}

Written in component form, the governing equations - we use the Euler
equation - for axi-symmetric problems are
\begin{displaymath}
  \matdiff{u'} - = -\frac{1}{\rho}\pdiff{p'}{r'};
  \ \matdiff{w'} = -\frac{1}{\rho}\pdiff{p'}{z'} - g,
  \ \left(\textrm{ where }\matdiff{}\equiv \pdiff{}{t'} + u'\pdiff{}{r'}+w'\pdiff{}{z'}\right).
\end{displaymath}
The continuity equation is expressed as
\begin{displaymath}
  \frac{1}{r'}\pdiff{}{r'}(r'u') + \pdiff{w'}{z'}=0.
\end{displaymath}
The boundary conditions are
\begin{displaymath}
  w'=\matdiff{h'} \ \mathrm{and} \ p'=\mathrm{Constant}\quad 
  \textrm{on }z'=h'(r',t'),
\end{displaymath}
and
\begin{displaymath}
w'=u'\diff{b'}{r'} \quad \textrm{on } z'=b'(r').
\end{displaymath}
The equations are quoted here in dimensional form (denoted by the
prime).  We now nondimensionalise these equations by introducing the
following scales: $h_0$, a typical undisturbed depth; $a$, a typical wave
amplitude; $\lambda$, a typical wave length; see
Fig(\ref{fig:config}).  The appropriate speed scale is $\sqrt{gh_0}$,
and this is used to define a suitable time scale; we now obtain the
familiar nondimensional equations
\begin{displaymath}
  u_t + \epsilon\left(uu_r+ wu_z \right)= -p_r;\ 
  \delta^2 \left\{ w_t + \epsilon\left(uw_r + ww_z \right) \right\}
  = -p_z,
\end{displaymath}
together with
\begin{displaymath}
  u_r + \frac{u}{r} + w_z=0;
\end{displaymath}
the upper boundary conditions become $p=\eta$, $w=\eta_t + \epsilon
u\eta_r$ on $z=1+\epsilon\eta$; the lower boundary condition is $w=
ub_r$ on $z=b(r)$.  In the above equations we have introduced
$p(r,z,t)$ to be the deviation from the undisturbed hydrostatic
pressure distribution, and the parameters are defined by
\begin{equation}
\epsilon=\frac{a}{h_0}\ \textrm{ and }\ \delta=\frac{h_0}{\lambda}.
\end{equation}

In addition to the above equations, we also have the equation for
global mass conservation.  If we consider the continuity equation
together with the boundary conditions, and if undisturbed conditions
exist far enough ahead of and behind the wave, then we can show that
\begin{displaymath}
  \intg_0^\infty \eta(r,t) r\d r =\mathrm{constant},
\end{displaymath}
which is the conservation of mass for the surface wave in cylindrical
geometry.  We will suppose that when the wave is initiated at a given
radius, it has a given profile; it therefore carries a known amount of
mass, the above constant.

These governing equations contain the familiar parameters for
water-wave theory; we now proceed with the conventional transformation
relevant in a far field(\cite{johnson80})
\begin{displaymath}
  \eta = \frac{\epsilon^3}{\delta^2}H \ ,\ 
  p = \frac{\epsilon^3}{\delta^2} P \ ,\ 
  u = \frac{\epsilon^3}{\delta^2}U \ , 
  r= \frac{\delta^2}{\epsilon^2}R \ ,\ 
  t = \frac{\delta^2}{\epsilon^2} \tau \ , \ 
  w=\frac{\epsilon^5}{\delta^4}W.
\end{displaymath}
This transformation accommodates the geometric decay of the surface
wave as $r\to\infty$.  At present the topology of the bottom, $b(r)$,
changes on the same scale as the wave evolves, so we introduce a
parameter $\alpha$ according to
\begin{displaymath}
b(r)=B(\alpha r),
\end{displaymath}
which will allow us to vary the scale on which the bottom changes
(\cite{johnson94}).  Introducing these into our governing equations we
obtain
\begin{eqnarray}
  U_\tau + \Delta \left(UU_R + WU_z \right) &=& -P_R;\label{govstart}\\
  \Delta \left\{ W_\tau + 
    \Delta \left( UW_R + WW_z \right)\right\} &=& -P_z;\\
  U_R + \frac{U}{R} + W_z &=&0,
\end{eqnarray}
with
\begin{eqnarray}
  P&=&H;\\
  W&=& H_\tau + \Delta UH_R,
\end{eqnarray}
both on $ z=1+ \Delta H$ (the rescaled free surface) ,and
\begin{equation}
  W=UB_R \quad \textrm{on }z=B\left(
    \frac{\delta^2\alpha}{\epsilon^2}R\right),\label{govend}
\end{equation}
where we have introduced
\begin{displaymath}
\Delta = \frac{\epsilon^4}{\delta^2},
\end{displaymath}
which will be the small parameter to be used in the problem, and will
be, therefore, the scale on which the nonlinear wave evolves.

There are then three cases of particular interest determined by the
size of $\alpha$ in relation to $\Delta$.  These are
\begin{displaymath}
  (1)\ \delta^2\alpha/\epsilon^2 = \Delta\sigma\,; \ 
  (2)\ \delta^2\alpha/\epsilon^2 = \Delta \, ; \ 
  (3)\ \delta^2\alpha/\epsilon^2 = \sigma , \ \Delta=\gamma\sigma,
\end{displaymath}
for $\Delta\to0$, $\sigma\to0$, $\gamma\to0$.  Case $(1)$ describes the
situation where the depth changes on a scale slower than that of the
evolution of the surface wave; it is this particular case which will
be dealt with in this presentation.

As an aside, case $(2)$ is where the depth changes on exactly the same
scale as the wave evolves; case $(3)$ is where the depth change is
slow, but faster than the evolution of the surface wave.

Before we proceed to investigate case $(1)$ - Slow Depth Change, we
must point out that in what follows, no assumption will be made
regarding the relationship between the two parameters $\Delta$ and
$\sigma$: they will be treated as independent small
parameters.

\section{Slow Depth Change}
To proceed we must introduce appropriate variables which are relevant
to the various propagation modes which exist in the problem.  The
initial surface profile, the Primary Wave which initiates the process,
propagates radially outwards; thus we need an outward characteristic
and (slow) scale to describe its evolution; these are
\begin{equation}
  \xi=\frac{1}{\Delta\sigma}\intg_1^Y F(y)\d y -\tau, \ 
  X=\frac{1}{\sigma}\intg_1^Y S(y)\d y , \ 
  Y=\Delta\sigma R,
\end{equation}
where the initial wave is centred about $Y=1$ i.e. $R=1/\Delta\sigma$
(a large radius).

The functions $F(Y)$ and $S(Y)$ are introduced to ensure that the
characteristics and slow scales are consistent with the depth
variation.  (Note: choosing $F=S=1$ recovers the classical
characteristics and slow scales for constant depth.)

We also anticipate that the propagation will initiate reflected
components travelling in the opposite direction, thus we also need a
suitable inward 'slow' characteristic
\begin{displaymath}
  \zeta=\intg_1^Y F(y)\d y +T , \ 
  T=\Delta\sigma\tau.
\end{displaymath}
Thus, any solution for the surface wave will be described (using a
multiple-scale approach) in terms of the variables $(\xi,\zeta,X,Y)$
and parameters $(\Delta,\sigma)$.  Introducing these new variables
into our governing equations (\ref{govstart})--(\ref{govend}), we
obtain
\begin{eqnarray*}
  \Delta\sigma U_\zeta - U_\xi + \Delta \left\{ FUU_\xi
    + \Delta SUU_X +\Delta\sigma FUU_\zeta + \Delta\sigma UU_Y 
    + WU_z \right\}\hspace{0.5cm} &\nonumber\\
  +FP_\xi + \Delta SP_X + \Delta\sigma FP_\zeta 
  + \Delta\sigma P_Y= &0;\label{eq:eq1}\\
  \Delta \left\{ \Delta\sigma W_\zeta - W_\xi + \Delta\left(FUW_\xi +
      \Delta SUW_X + \Delta\sigma FUW_\zeta 
      +\right.\right.\hspace{0.5cm}&\nonumber\\
  \left.\left. \Delta\sigma UW_Y + WW_z \right)\right\}+P_z=&0;\\
  FU_\xi + \Delta SU_X + \Delta\sigma FU_\zeta +\Delta\sigma U_Y 
  + \frac{\Delta\sigma U}{Y} + W_z =&0\label{eq:eq3},
\end{eqnarray*}
with boundary conditions (on $z=1+\Delta H$)
\begin{eqnarray*}
  P&=&H;\label{bc:bc1} \\
  W&=& \Delta\sigma H_\zeta - H_\xi + \Delta U \left\{ FH_\xi +
    \Delta SH_X + \Delta\sigma FH_\zeta 
    + \Delta\sigma H_Y \right\},\label{bc:bc2}
\end{eqnarray*}
and
\begin{displaymath}
W=-\Delta\sigma UD'(Y) \quad \textrm{ on }z=1-D(Y).
\end{displaymath}
Here we have introduced $D(Y)$, defined as
\begin{equation}
  D(Y)=1-B(Y),
\end{equation}
to be the local depth, a more convenient function than $B(Y)$.

We now seek a solution to these equations in the form of a double
asymptotic expansion
\begin{displaymath}
  Q\sim\sum_{n=0}^\infty \sum_{m=0}^\infty \Delta^n\sigma^mQ_{nm}, 
  \quad \Delta\to0, \ \sigma\to0,
\end{displaymath}
where $Q$ represents each of $U,W,P,H$.

\section{Results}
We proceed by constructing systems of equations at each order,
$\Delta^n\sigma^m$, and imposing conditions (as appropriate) which
ensure that the asymptotic expansions remain uniformly valid as
$|\xi|\to\infty, |\zeta|\to\infty$ .  We present only the main
results that occur at each order (using results from previous orders,
where relevant):

{$\Delta^0\sigma^0$:}
\begin{equation}\label{eq:fy}
  F(Y)=\frac{1}{\sqrt{D(Y)}}.
\end{equation}
{$\Delta^1\sigma^0$:}
\begin{equation}\label{eq:primary}
  2 \sqrt{D}S H_{00X} + \frac{3}{D}H_{00}H_{00\xi} 
  + \frac{D}{3}H_{00\xi\xi\xi}=0.
\end{equation}
{$\Delta^1\sigma^1$:}
\begin{equation}\label{eq:outshelf}
  2\sqrt{D}SH_{01X} + \frac{3}{D}(H_{00}H_{01})_{\xi} 
  + \frac{D}{3}H_{01\xi\xi\xi}= 
  -2\frac{D^{1/4}}{Y^{1/2}}\left( D^{1/4}Y^{1/2}H_{00}\right)_Y.
\end{equation}
{$\Delta^2\sigma^2$:}
\begin{equation}\label{eq:inshelf}
  \mathcal{U}_{11\zeta}+\frac{\Lambda_{11\zeta}}{\sqrt{D}} 
  + \Lambda_{11Y}=0; 
  \qquad \sqrt{D}\mathcal{U}_{11\zeta} + \Lambda_{11\zeta} 
  + \frac{(DY\mathcal{U}_{11})_Y}{Y}=0,
\end{equation}
and
\begin{equation}\label{eq:newoutshelf}
  2D^{1/4}Y^{1/2}(D^{1/4}Y^{1/2}h_{11\xi})_Y = -(DYH_{00Y})_Y,
\end{equation}
where we have introduced
\begin{eqnarray*}
U_{11} &=& u_{11}(\xi,X,Y) + \mathcal{U}_{11}(\zeta,X,Y);\\
H_{11} &=& h_{11}(\xi,X,Y) + \Lambda_{11}(\zeta,X,Y).
\end{eqnarray*}
Equations (\ref{eq:primary}) and (\ref{eq:outshelf}), when suitably
combined, imply
\begin{displaymath}
  Y\sqrt{D}\intg_{-\infty}^{\infty} H_{00}^2 \d Y = \mathrm{Constant},
\end{displaymath}
the conservation of 'momentum' for the cylindrical KdV equation.

\begin{figure}
  \begin{center}
    \includegraphics[height=7cm,width=0.8\textwidth]{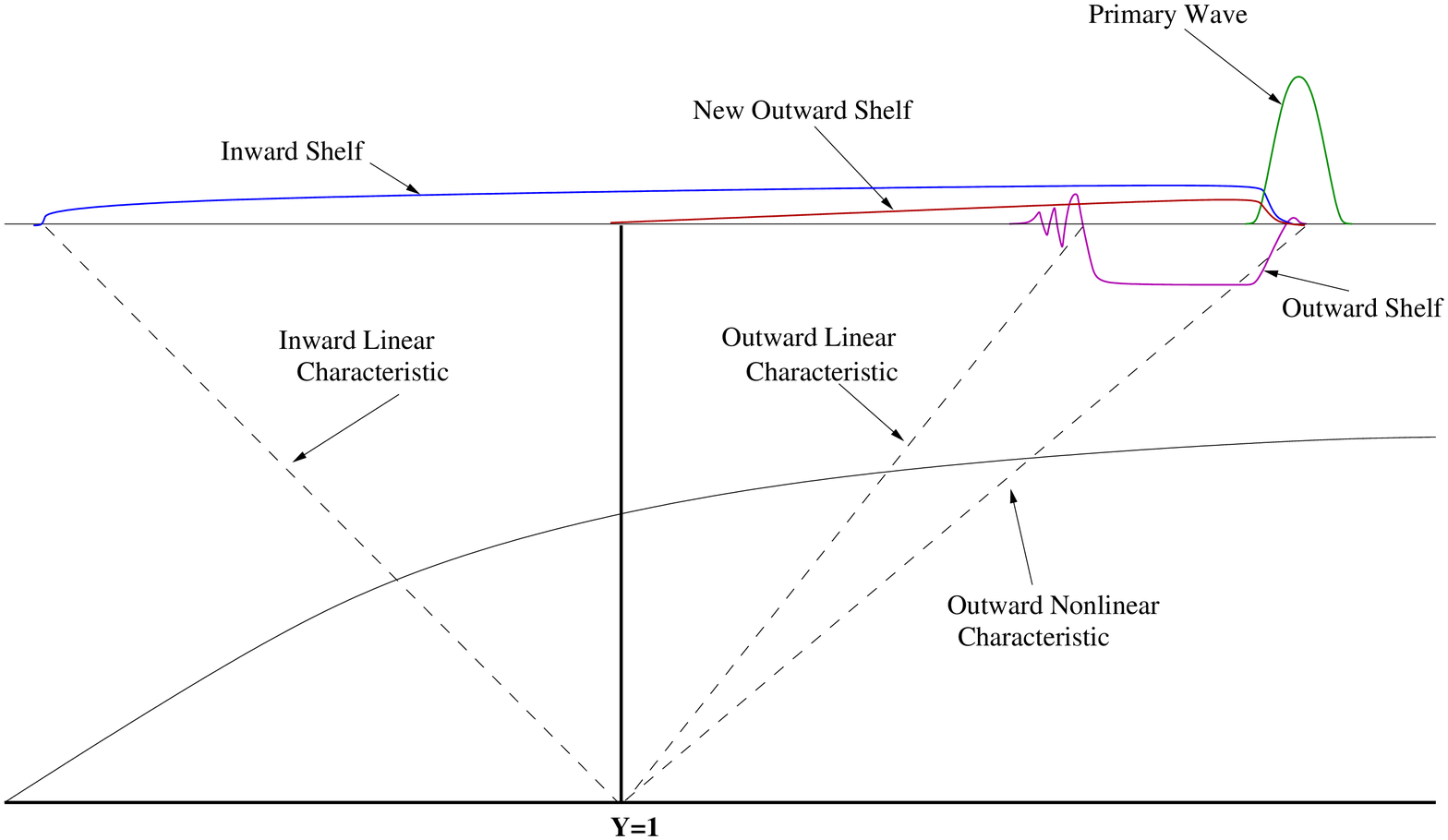}
  \end{center}
\caption{Schematic of Components in problem}\label{fig:components}
\end{figure}

In passing we observe the similarities between the above equations for
the radially symmetric problem and those for the plane-wave
problem \cite{johnson94}; it is this similarity that suggests a
similar approach to finding a solution.  Note that, by virtue of the
large-radius scaling, the geometric contribution in the cylindrical
coordinates first appears in (\ref{eq:outshelf}), leaving
(\ref{eq:primary}) as the classical KdV equation.  The various
components described by equations
(\ref{eq:primary})--(\ref{eq:newoutshelf}) are represented schematically in Fig(\ref{fig:components}); see the descriptions that
follow, where we give a brief outline of what can be deduced at each
order.

\subsection{Primary Wave}
Equation (\ref{eq:primary}) is a variable coefficient KdV equation;
thus our $\sech^2$ initial profile is an exact solution for general
$S(Y)$, however, in our particular problem we require
$S(Y)=D^{-5/2}(Y) Y^{-2/3}$, in order to remove nonuniformities; this,
along with the requirements of conservation of momentum, yields a
solution for the primary wave as
\begin{equation}
  H_{00} = \frac{A_0}{DY^{2/3}} \sech^2
  \left(\frac{\sqrt{3A_0}}{2D^{3/2}Y^{1/3}} \left(\xi-cX\right)\right),
\end{equation}
where $A_0$ is the initial amplitude (at $Y=1$) of the wave.

\subsection{Outward 'Shelf'}
Now let us consider (\ref{eq:outshelf}); we anticipate that, as this
equation has distinct similarities to that of the plane-wave case, the
solution to this equation will exhibit shelf-like properties.

Let us introduce $\phi=\xi-cX$, to be a characteristic moving with the
primary wave (with $c$ the constant speed in $(\xi,X)$ space), and let
$\phi=0$ be the peak.  To demonstrate that the component $H_{01}$ does
not approach zero near the tail of the primary wave, we construct
an integral across a neighbourhood of $\phi=0$.  This shows that
\begin{displaymath}
  Z_{01}(T)=-\frac{m_0 \bar{D}^{7/4}\bar{Y}^{1/6}}{c}(\bar{D}^{3/4}
  \bar{Y}^{1/6})_T,
\end{displaymath}
is the amplitude of $H_{01}$ directly behind the primary wave.  Here
we have defined $Y=\bar{Y}(T)$ to be $\phi=0$, describing the path of
the primary wave.

From (\ref{eq:outshelf}) we can also obtain a complete description of
this wave component behind the primary wave.  When we note that
$H_{00}$ is exponentially small in its tails, we find that defining
$H_{01}=D^{-1/4}Y^{-1/2}G_{01}$, then
\begin{displaymath}
  G_{01} \sim \mathcal{G}' \left( \sigma \phi + c \intg_0^T 
    \bar{D}^{-2}\bar{Y}^{-2/3} dT \right),
\end{displaymath}
where
\begin{displaymath}
  \mathcal{G}(\mathcal{T}) = -m_0 \bar{Y}^{1/6}\bar{D}^{3/4} 
  +\mathcal{A} \qquad \bar{Y}=\bar{Y}(\mathcal{T}),
  \bar{D}=\bar{D}(\mathcal{T}),
\end{displaymath}
and
\begin{displaymath}
  \mathcal{T} =  c \intg_0^T \bar{D}^{-2}\bar{Y}^{-2/3} dT.
\end{displaymath}
At the front (outside end) there exists a transition region (a region
where transition back to undisturbed conditions occurs) around
$\phi=0$; similarly, at the rear (inside end), a transition region
exists around $\xi=0$.  The detailed structure of these transitions
can be obtained from (\ref{eq:outshelf}) by considering the relevant
asymptotic regions.  They turn out to be, at the front,
\begin{displaymath}
H_{01}= D^{9/4}Y^{1/6}\left(D^{3/4}Y^{1/6}\right)_Y L \ ,
\end{displaymath}
where\begin{eqnarray*}
  L=& \frac{8}{3\sqrt{3}} {A_0}^{-5/2} \left\{ \frac{3{A_0}^2}{2} (\tanh  
    \theta -1) + \left(3{A_0}^2\theta + \frac{9{A_0}^2}{2}\right)\sech^2 
    \theta\right.\\
  &\hspace{3cm} \left. - \tanh\theta\sech^2\theta 
    \left(\frac{9{A_0}^2 \theta}{2} + \frac{\hat{s}}{2} + 3{A_0}^2 
      \theta^2 \right) \right\},
\end{eqnarray*}
with
\begin{displaymath}
  \theta = \frac{\sqrt{3A_0}}{2 D^{3/2}Y^{1/3}} 
  \left(\xi - cX\right), 
  \qquad\hat{s}=\mathrm{constant};
\end{displaymath}
and at the rear
\begin{displaymath}
  H_{01} = D^{-1/4}Y^{-1/2}\mathcal{G}'(0)
  \left\{1- \int^\infty_{x} Ai(u)du \right\};
  \quad x=\frac{\xi}{DY^{2/9}}\left(\frac{2}{X}\right)^{1/3}.
\end{displaymath}

\begin{figure}
 \begin{center}
    \includegraphics[height=0.7\textwidth,angle=270]{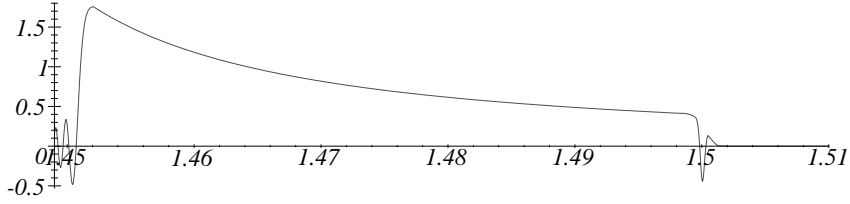}
\end{center}
\caption{An example of the outward shelf 
    behind the primary wave ($H_{01}$ against $Y$) 
    using $A_0=1$, $\sigma=0.01$, $\Delta=0.1$ and $\bar{Y}=1.5$.}
  \label{fig:outshelf}
\end{figure}

In Fig(\ref{fig:outshelf}) we show a solution for the outward shelf
combining the above expressions.  We have made use of a simple depth
profile, as an example ($D(Y)=Y^{-2}$), and have made necessary
choices for the parameters and constants, which allow the three
asymptotic solutions to be suitably joined at this order of
approximation.

\subsection{Mass Conservation}
As we have already mentioned, the specific nature of this problem
leads to an underlying requirement for global conservation of mass.
After the application of all of the relevant scalings we obtain the
mass conservation condition as
\begin{eqnarray*}
  \intg_0^\infty H(R,\tau) R\d R 
  &=& \intg_0^\infty (H_{00}+ \sigma H_{01} 
  + \Delta H_{10}+ \Delta\sigma H_{11}+\ldots)R \d R\\ 
  &=&\mathrm{Constant}.
\end{eqnarray*}
Now let us consider the mass carried by the primary wave; this turns
out to be
\begin{displaymath}
  \intg_0^\infty H_{00}R\d R \sim 
  \bar{D}^{1/2}\bar{Y}\intg_{-\infty}^{\infty} H_{00}\d\phi = 
  {m_0\bar{D}}{\bar{Y}^{2/3}},
\end{displaymath}
a quantity which is clearly not constant.  Thus the resulting
difference from the initial O(1) mass, $m_0$, must be carried by other
wave components.

We can calculate the mass carried by the Outward Shelf as
\begin{displaymath}
  {\sigma}\intg_0^\infty H_{01}(\xi,X,R)R\d R \sim 
  m_0\left({\bar{D}^{1/4}}{\bar{Y}^{1/2}} - 
    {\bar{D}}{\bar{Y}^{2/3}}\right),
\end{displaymath}
which satisfies the condition that the mass carried by this component
is zero at $Y=1$.  We can see that the total mass carried by these
two components is
\begin{displaymath}
  \intg_0^\infty (H_{00}+\sigma H_{01})R\d R=  
  {m_0\bar{D}^{1/4}}{\bar{Y}^{1/2}} + O(\Delta),
\end{displaymath}
Thus there must be other components carrying $O(1)$ mass.

\subsection{Re-reflected Shelf}

Let us now investigate the mass carried by the outward moving
component of $H_{11}$. (We have written the non-local contributions to
$H_{11}$ as $h_{11}+\Lambda_{11}$; see
(\ref{eq:inshelf}),(\ref{eq:newoutshelf}).)  This component is
governed by (\ref{eq:newoutshelf}) which can be integrated to yield
\begin{displaymath}
  h_{11\xi}=-\frac{1}{2}D^{-1/4}Y^{-1/2}\intg_1^Y 
  D^{-1/4}Y^{-1/2}\pdiff{}{Y}\left\{DYH_{00Y}\right\}dY.
\end{displaymath}
For general depth profiles we are unable to obtain an explicit
description for this wave component. However, the use of a specific
form for the depth change, namely
\begin{equation}
  D(Y)=\frac{(\beta+\gamma Y^2)^{4/3}}{Y^2}, \label{eq:depth}
\end{equation}
enables us to completely solve this equation and thereby elucidate
some properties of this component.

Calculating the mass carried by this component, $M_0$, we now find that
\begin{eqnarray*}
  M_0&=&{m_0} \intg_1^{\bar{Y}} \left[ 
    \frac{2\gamma (\beta+\gamma Y^2)^{2/3}}{Y^{4/3}} 
    - \frac{4(\beta+\gamma Y^2)^{4/3}}{3Y^{7/3}}\right. \\
    &&\hspace{3cm}+\left. \left(\frac{4}{3}-2\gamma\right)Y(\beta
    +\gamma Y^2)^{-1/3}\right] dY.
\end{eqnarray*}

\subsection{Reflected Shelf}

This component is governed by the pair of linear homogeneous equations
(\ref{eq:inshelf}).  In order that we have conserved $O(1)$ mass, this
component must be carrying the mass
\begin{displaymath}
  m_0 - {m_0 \bar{D}^{1/4}}{\bar{Y}^{1/2}}-M_0,
\end{displaymath}
and this requirement uniquely determines $\Lambda_{11}$.  The details
are involved, but readily accessible (at least for suitable $D(Y)$).

The full solutions for both the reflected and the re-reflected
shelves, for our specific choice for the depth profile
(\ref{eq:depth}), are not presented here; but they can be seen, along
with the description of the transitions regions for the re-reflected
shelf, in \cite{killen99}.

\section{Conclusions}

In this work we have presented a description of the wave components in
cylindrical geometry that carry $O(1)$ mass.  It has, however, not
been possible to present all the details for general depth profiles
(which would have been preferable).  Nevertheless, we have shown the
development of the method of solution, and from this can obtain some
particular results.

Thus, for example, we find that if the depth profile takes certain
(very simple) profiles, e.g. $D(Y)\propto Y^{-2/9}$, then one of the
wave components, the new outward (re-reflected) shelf, will be absent.

One fundamental question which still has to be answered, concerns the
behaviour of the inward (reflected) wave component.  It is obvious, by
the very nature of the problem, that any component propagating inwards
will eventually reach the centre.  There will undoubtably be some form
of singularity at the centre, but there may be some choice for the
depth profile in the neighbourhood of $r=0$ which will result in a
particularly simple structure.  The details of this investigation are
still being developed, and may provide an impetus for further study.


\end{document}